\documentclass[10pt,doublecolumn]{IEEEtran}
\usepackage{amsmath,amsthm,amsfonts,amssymb,bm,mathrsfs}
\usepackage{graphicx,subfigure}
\usepackage{epstopdf}
\usepackage{cite}
\usepackage{algorithm}
\usepackage{algorithmic}
\usepackage{multirow}
\usepackage[]{caption2}

\graphicspath{{fig/}}
\theoremstyle{remark}

\theoremstyle{plain}

\title{Deep Learning-based Intelligent Dual Connectivity for Mobility Management in Dense Network}
\author{\IEEEauthorblockN{Chujie Wang$^1$, Zhifeng Zhao$^1$, Qi Sun$^2$, Honggang Zhang$^1$}\\
	\IEEEauthorblockA{$^1$College of ISEE, Zhejiang University, Zheda Road 38, Hangzhou 310027, China\\
	$^2$China Mobile Research Institute, Beijing 100053, China}\\
	\IEEEauthorblockN{Emails:$\{$21731125, zhaozf, honggangzhang$\}$@zju.edu.cn, sunqiyjy@chinamobile.com}}
\begin{document}
    \maketitle
	\begin{abstract}
		Ultra-dense network deployment has been proposed as a key technique for achieving capacity goals in the fifth-generation (5G) mobile communication system. However, the deployment of smaller cells inevitably leads to more frequent handovers, thus making mobility management more challenging and reducing the capacity gains offered by the dense network deployment. In order to fully reap the gains for mobile users in such a network environment, we propose an intelligent dual connectivity mechanism for mobility management through deep learning-based mobility prediction. We first use LSTM (Long Short Term Memory) algorithm, one of deep learning algorithms, to learn every user equipment's (UE's) mobility pattern from its historical trajectories and predict its movement trends in the future. Based on the corresponding prediction results, the network will judge whether a handover is required for the UE. For the handover case, a dual connection will be established for the related UE. Thus, the UE can get the radio signal from two base stations in the handover process. Simulation results verify that the proposed intelligent dual connectivity mechanism can significantly improve the quality of service of mobile users in the handover process while guaranteeing the network energy efficiency.
	\end{abstract}
	\section{Introduction}
	The fifth-generation (5G) mobile communication system is expected to offer 1000-fold capacity increase compared with the current 4G deployments, aiming at providing higher data rates and lower end-to-end delay while supporting high-mobility users\cite{andrews2014will}. To this end, ultra-dense network deployment has been proposed as a key technology for achieving the capacity goal\cite{kamel2016ultra}. Compared with the conventional networks, ultra-dense networks consist of a large number of small cells, so as to offer higher throughput for static users\cite{yunas2015spectral}, while simultaneously bring challenging issues for moving users\cite{wu2016survey}. Within ultra-dense networks, user equipment(UE) in the mobility will frequently cross cells, thus attaching to different base stations (BSs) to maintain the connection with the serving BS providing strongest signal strength. In the formal terminology, the whole procedure from one attached BS to another is called a handover (HO).
	
	Traditional HO adopts a passive trigger-based strategy, making the mobile network no prior preparation. This posterior handover incurs negative impact on both the user side and the network side. On the user side, since an HO involves intensive signaling interactions between the UE, serving BS, target BS, and core networks, it usually leads to large delay and obvious throughput reduction. On the network side, due to its realistic loads, one target BS may reject the access request from handover UEs in its busy period. Therefore, the impact from both sides will decrease the quality of service and possibly make the network densification in vain. Hence, it is critical to design novel mobility management techniques.
	
	There has been a substantial body of researches toward solving this issue. Recently, it has been proved that human mobility is predictable to some extent. Song et al.\cite{song2010limits} quantitatively analyze the regularity of the moving trajectory using the information theory. They find that the user mobility contains 93\% potential predictability by calculating the conditional entropy of the position sequence of one user's history motion, which illustrates the feasibility of learning users' mobility patterns through their history trajectory information. The authors in \cite{becvar2010adaptive} propose an adaptive HO Hysteresis Margin (HHM) method to reduce the number of HOs. They use the predefined Reference Signal Received Quality (RSRQ) threshold and path-loss factor to adjust the HHM. The work in \cite{zhang2010novel} studies the HO process in two-tier macrocell-femtocell networks and proposes a representative cost function-based algorithm, where the function considers a weighted sum of parameters related to UE speed, cell load and the number of user connections. Then, the cost function results are integrated into a typical  Received Signal Strength (RSS)-based procedure as an adaptive factor to change the HHM. The work in \cite{tsai2016using} introduces the fuzzy logic to analyze and adjust the HO parameters. However, due to the flaws of fuzzy logic itself, there are some inaccurate variable descriptions in the calculation process, making the results somewhat subjective. In a word, the aforementioned threshold-tuning based algorithms lack systematic methodologies to optimize the HO-related parameters, which may hinder practical applications. Another type of HO optimization strategies is based on the user's mobility status. The authors in \cite{guidolin2016context} propose a time-to-trigger parameter selection policy by assuming that the user's trajectory is already known and trying to balance the handover failure probability and the ping-pong effect. The authors in \cite{lee2015time} put forward an approach for the prediction of user's movement direction and residence time in the target cell to reduce unnecessary HOs. However, the prediction result in \cite{lee2015time} is obtained based on the current value only rather than the complete history information, which decreases the prediction accuracy. In summary, the results in the literature poorly leverage the merits in users' historical mobility and lack a more comprehensive combination between the prediction results and the HO procedure.
	
	In order to improve the HO performance of mobile users in ultra-dense networks, this paper proposes an intelligent dual connectivity mechanism for mobility management by taking advantage of deep learning-based mobility prediction\cite{han2017big} \cite{chih2017big}. On one hand, we use LSTM (Long Short Term Memory) scheme to fully learn the user's mobile patterns from its history trajectories and predict its movement trends in the future. On the other hand, based on the mobility prediction results for a UE, the network will start a dual connection between the UE and two BSs (e.g., one serving BS and one BS for potential HO). Therefore, some preparation could be made in advance and the practical HO could be executed more efficiently.
	
	This paper is organized as follows. In Section II, we describe the proposed intelligent dual connectivity mechanism in detail. Section III introduces the LSTM algorithm which we used for mobility prediction. Section IV presents the simulation results from different aspects. Section V concludes this paper.
	
	\section{Dual Connectivity Method}		
\begin{figure}[htbp]
		\begin{center}
			\includegraphics[width=0.35\textwidth]{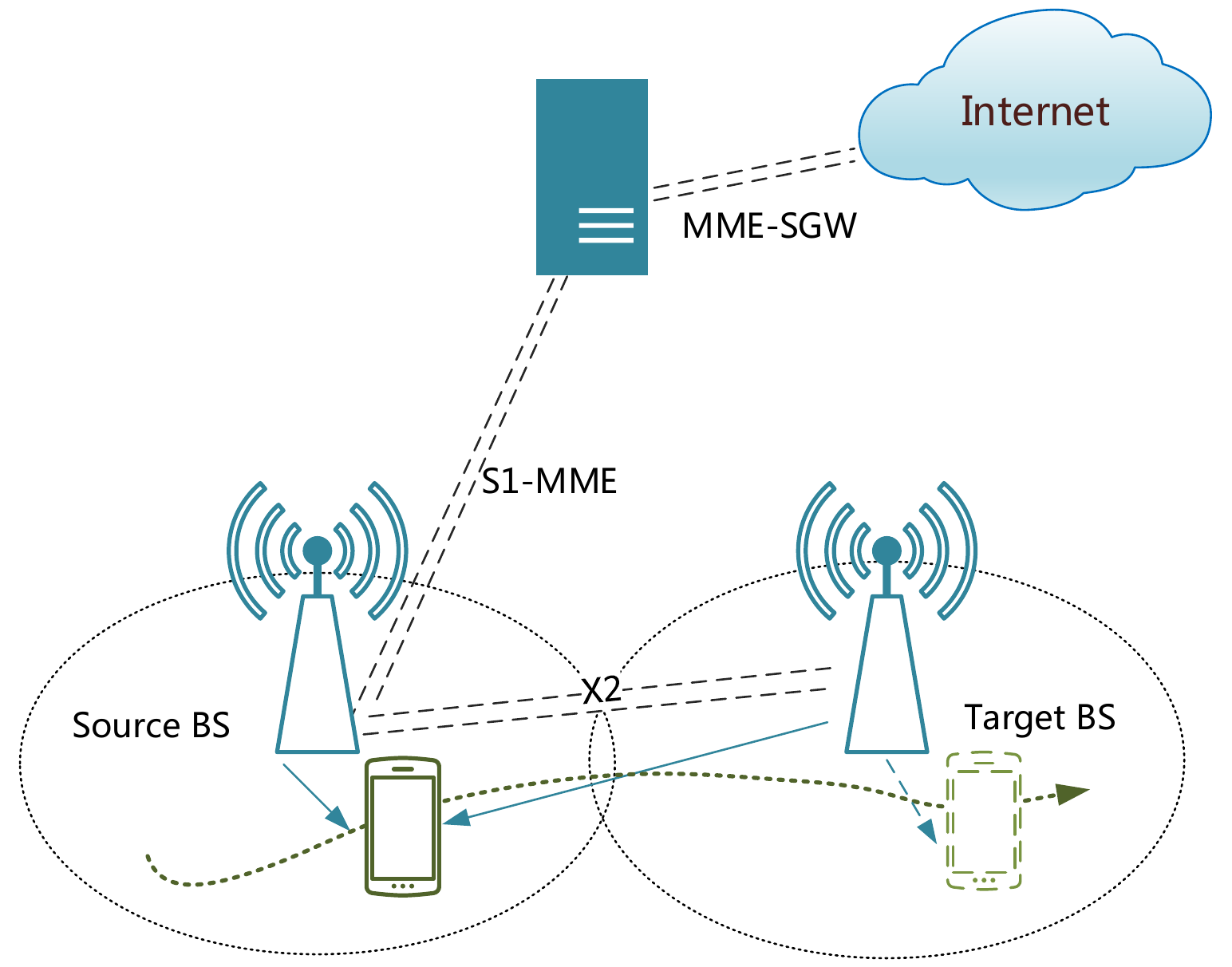}
			\setlength\abovecaptionskip{5pt}
			\setlength\belowcaptionskip{-5pt}
			\captionsetup{font={small}}
			\caption{A simplified architecture of dual connectivity}
			\label{fig:1}
		\end{center}
	\end{figure}

\begin{figure}[htbp]
		\begin{center}
			\includegraphics[width=0.47\textwidth]{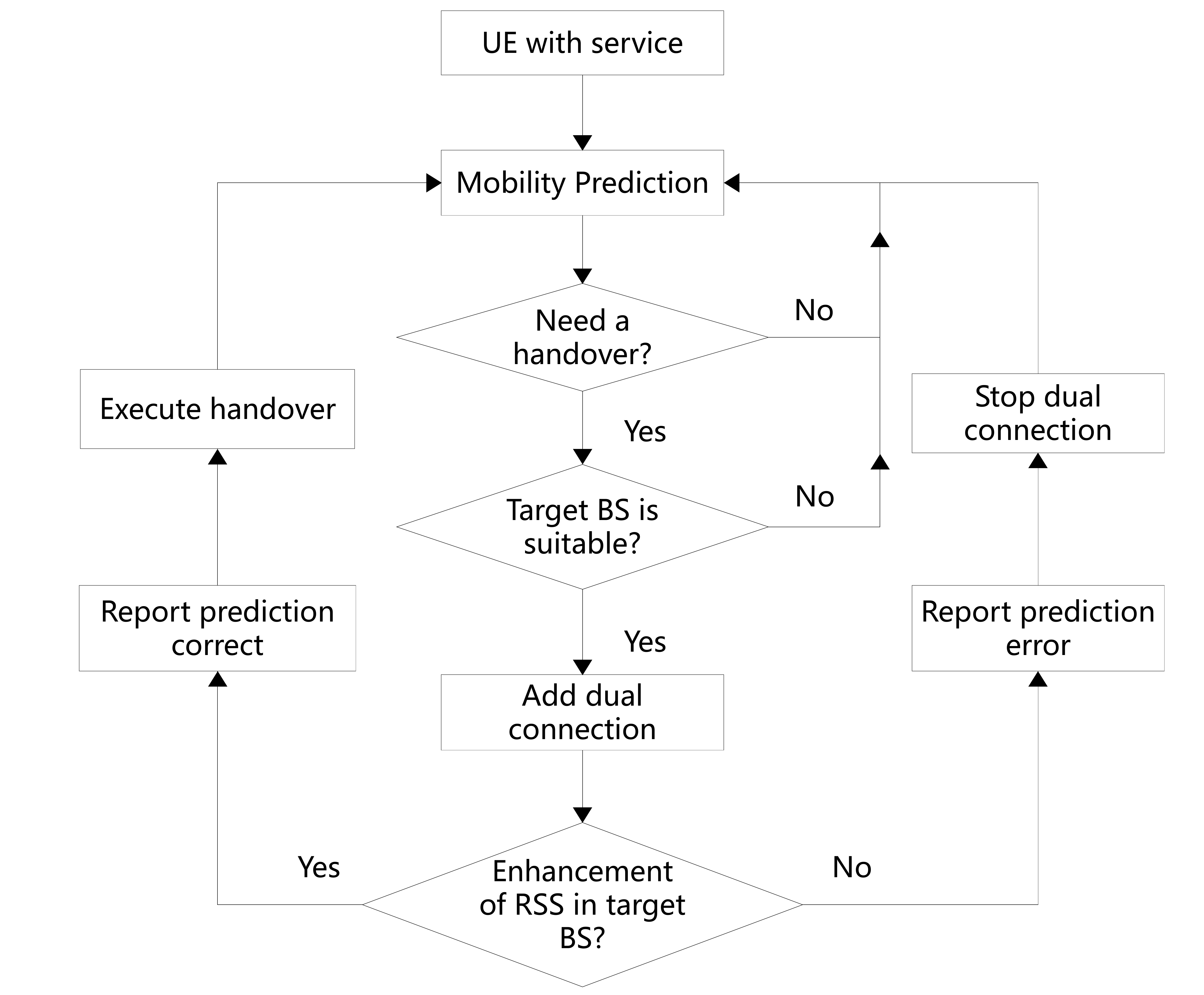}
			\setlength\abovecaptionskip{5pt}
			\setlength\belowcaptionskip{-5pt}
			\captionsetup{font={small}}
			\caption{The concrete process of dual connectivity}
			\label{fig:2}
		\end{center}
	\end{figure}

	A simplified schematic diagram of the proposed dual connectivity architecture is illustrated in Fig. \ref{fig:1}. The moving UE is simultaneously connected to its source BS (serving BS) and target BS. The serving BS contains both user-plane and control-plane connection, while the target BS only transmits user-plane messages. Therefore, for each dual-connected UE, there is only a single connection toward the mobility management entity (MME), through the S1 interface that links the serving BS to the core network (CN). However, the two BSs can exchange radio resource control (RRC) configuration messages via X2 link, which may be a wired or wireless backhaul. Fig. \ref{fig:2} describes the specific process of the proposed dual connectivity mechanism.
	
	Basically, the user mobility model can be trained in advance from the user's history position data. Also when a mobile UE is under service, the network (e.g., the mobility management entity, MME) can make a prediction about its future position at a certain moment based on the historical records. The training and prediction process can be accomplished by LSTM-based deep learning scheme, which will be explained with details in the next section. According to the spatial distribution of BSs and the prediction result, the MME evaluates future RSS of the UE and determine its best serving BS at that time, which will lead to two situations. If there is no change of the best serving BS for the UE, indicating that no HO is needed, then the MME takes no action. Otherwise, the MME will check the load status of the target BS and judge whether or not to trigger a dual connection. If yes, the MME informs the source BS to prepare for the dual connectivity.
	
	However, due to the strong randomness of users' mobility, the prediction cannot be completely accurate sometimes. In order to avoid the energy loss caused by unnecessary dual connection, we establish a protection mechanism to deal with any wrong prediction. Specifically, the dual-connected UE continuously monitors the RSS of its source BS and target BS. Thus, the UE could judge whether it is heading to the dual-connected target cell and inform the MME of its measurement result. Meanwhile, the MME will further re-train its prediction algorithm based on the UE report. When the conventional HO conditions are satisfied, the UE will execute the HO to its target cell, after which the connection with the original serving BS will be released and the target BS will become the UE's new serving BS. Then the UE maintains this strongest connection with the current serving BS. On the contrary, if the RSS from the target BS stays at a low level, the prediction is likely to be wrong, then the UE reports the error message to network and deletes the dual connection with the predicted BS immediately.
	
	\section{Deep Learning with LSTM Scheme}
	Within the framework of deep learning, recurrent neural network (RNN) has been widely used in time-series prediction. However, it's difficult for conventional RNN to make an accurate prediction when it needs to remember long-term memories. Therefore, the LSTM has been proposed to solve the problem of long-dependency. In particular, the LSTM network is composed of multiple copies of basic memory blocks and each memory block contains a memory cell and three types of gates (input gate, output gate, and forget gate). The memory cell is the key component of LSTM and responsible for information transfer at different time steps. Meanwhile, the three gates, each of which contains a sigmoid layer to optionally pass information through, are responsible for protecting and controlling the cell state. Specifically, the input gate controls which part of the input will be utilized to update the cell state. Similarly, the forget gate controls which part of the old cell state will be thrown away, while the output gate determines which part of the new cell state will be output.
	\begin{figure}[htbp]
		\begin{center}
			\includegraphics[width=0.38\textwidth]{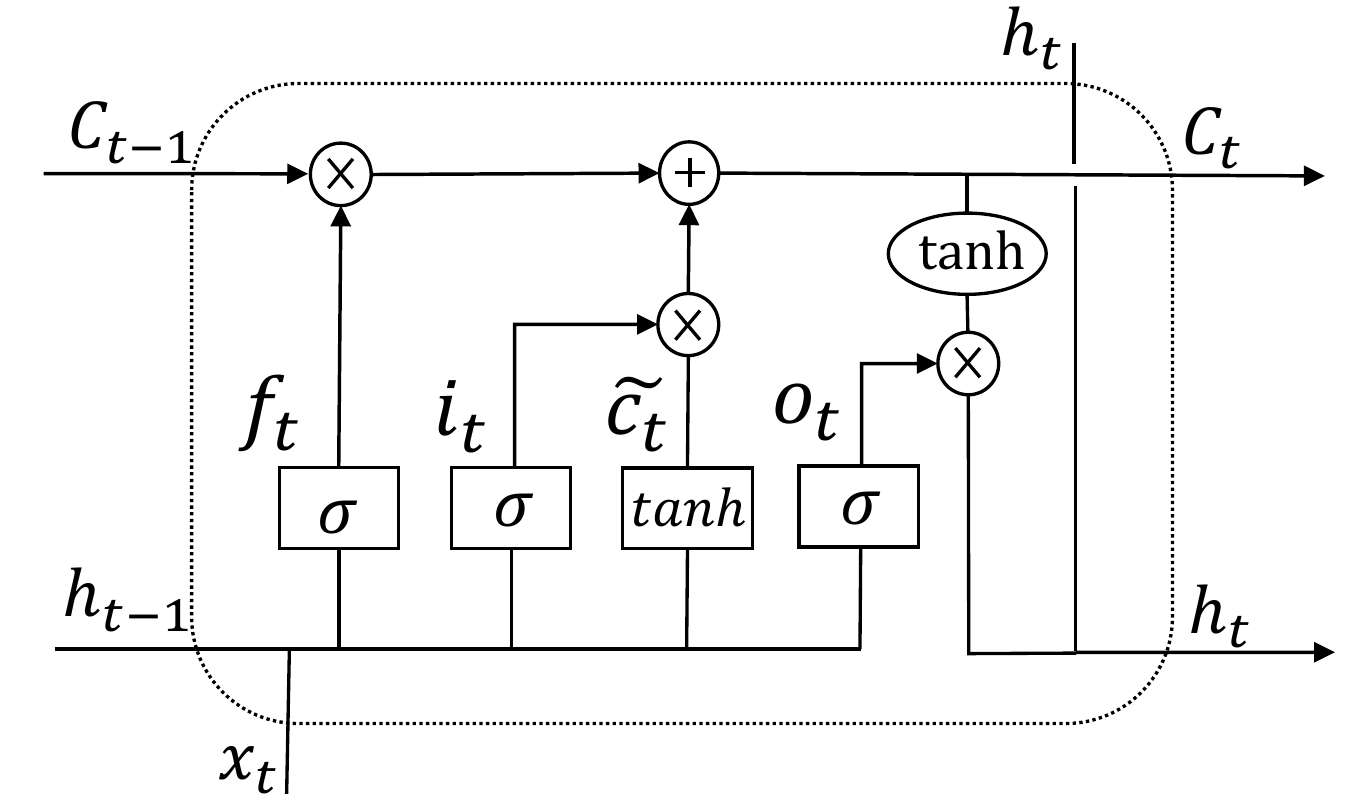}
			\setlength\abovecaptionskip{5pt}
			\setlength\belowcaptionskip{-5pt}
			\captionsetup{font={small}}
			\caption{LSTM memory block architecture}
			\label{fig:3}
		\end{center}
	\end{figure}
	
	As shown in Fig. \ref{fig:3}, for the memory block at time step $t$, we use $f_t$, $i_t$, and $o_t$ to represent the forget, input and output gates respectively. Assume that $x_t$ and $h_t$ represent the input and output at the current time step, $h_{t-1}$ is the output at the previous time step, $\sigma$ represents the sigmoid activation function, then the values of the three gates can be calculated as:
	\begin{equation}
	f_t=\sigma (W_f\cdot[h_{t-1},x_t]+b_f)
	\end{equation}
	\begin{equation}
	i_t=\sigma(W_i\cdot[h_{t-1},x_t]+b_i)
	\end{equation}
	\begin{equation}
	o_t=\sigma(W_o\cdot[h_{t-1},x_t]+b_o)
	\end{equation}
	where $W_f$, $W_i$, $W_o$, $b_f$, $b_i$, $b_o$ are corresponding weight matrices and variable biases of the three gates.
	
	Then the process of updating information through the gate structure can be divided into three steps. Firstly, multiplying the value of forget gate $f_t$ by the old cell state $C_{t-1}$ decides which part of the previous cell state $C_{t-1}$ should be thrown away. Then, let us update the information in the cell state by multiplying the value of input gate $i_t$ by the new candidate memory cell value $\tilde{C_t}$:
	\begin{equation}
		\tilde{C_t}=tanh(W_c\cdot[h_{t-1},x_t]+b_c)
	\end{equation}	
	\begin{equation}
		C_t=f_t*C_{t-1}+i_t*\tilde{C_t}
	\end{equation}
	where $W_c$, $b_c$ represent the weight matrix and variable bias of the memory cell. Finally, multiplying the output gate $o_t$ by the updated cell state $C_t$ through a tanh function leads to the output value $h_t$:
	\begin{equation}
		h_t = o_t*tanh(C_t)
	\end{equation}
	The output $h_t$ and cell state value $C_t$ will be passed to the memory block at the next time step.
	\begin{figure}[htbp]
		\begin{center}
			\includegraphics[width=0.40\textwidth]{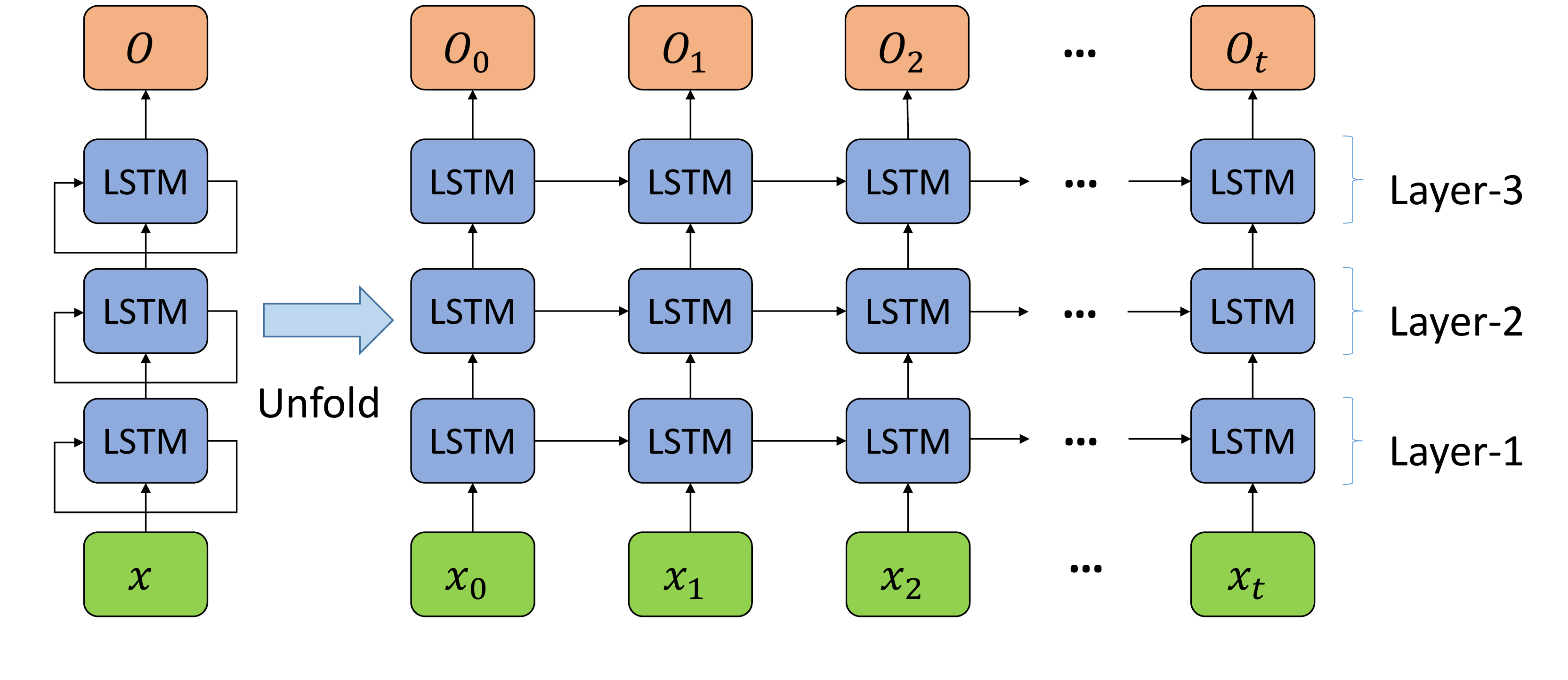}
			\setlength\abovecaptionskip{5pt}
			\setlength\belowcaptionskip{-5pt}
			\captionsetup{font={small}}
			\caption{The unfold structure of a three-layer LSTM neural network which we used in our work}
			\label{fig:4}
		\end{center}
	\end{figure}
	
	A deep recurrent neural network can be formed by stacking multiple hidden layers. Fig. 4 shows the unfold structure of a three-layer LSTM network, which we use for the task of mobility prediction in our work. The first LSTM layer takes a two-dimension position sequence as input and feeds the output to the second layer, which in turn feeds its output to the last LSTM layer to get the final result at each time step.
	
	\section{Performance Evaluation and Simulation Results}
	\subsection{Dataset and Simulation Environment}
\begin{figure}[htbp]
		\begin{center}
			\includegraphics[width=0.34\textwidth]{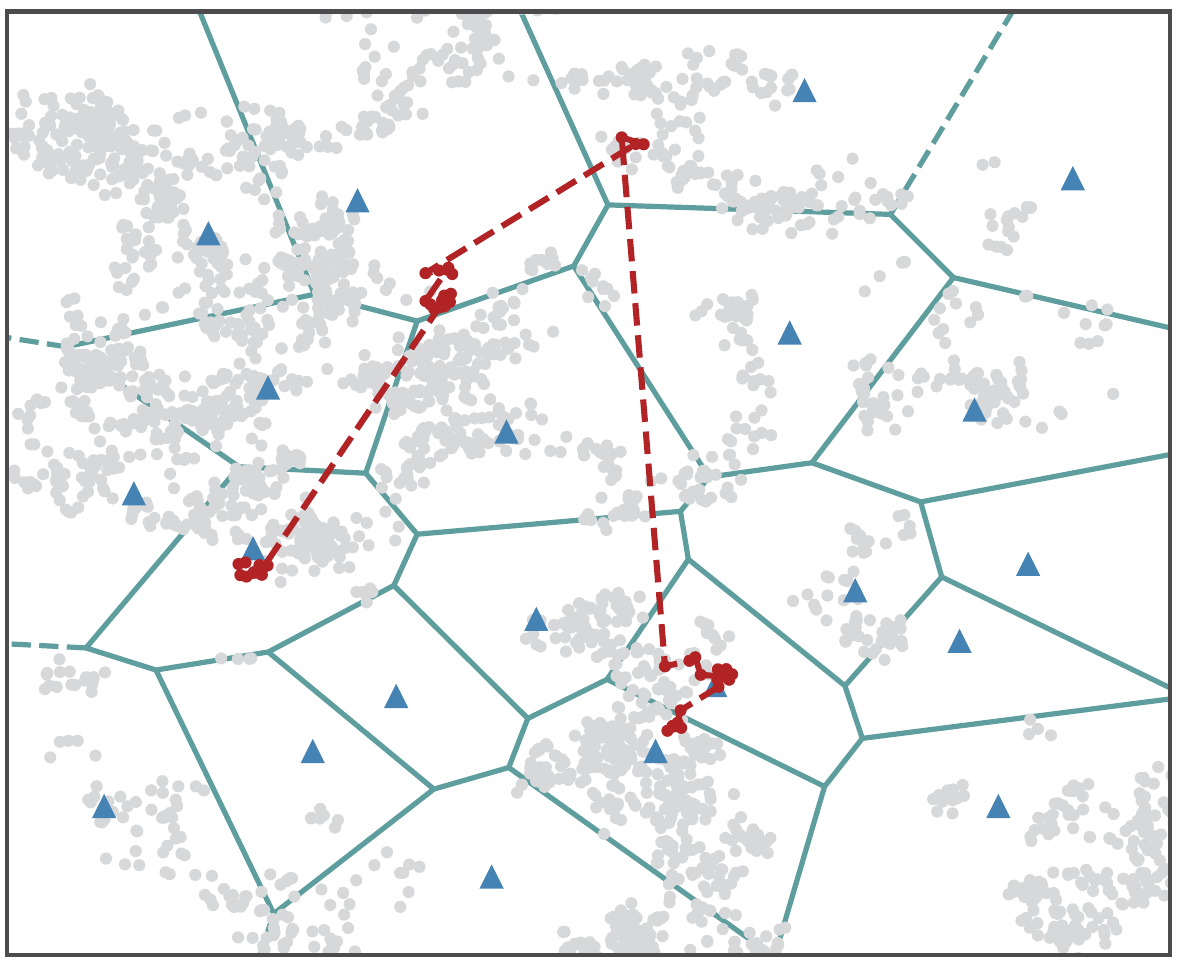}
			\setlength\abovecaptionskip{5pt}
			\setlength\belowcaptionskip{-5pt}
			\captionsetup{font={small}}
			\caption{The simulation area(4000m*4000m) with cell density $\lambda=1$ and the trajectory of a single user in a day. The red dotted line describes the trajectory of the user. The gray dots and blue triangles represent hotspots and BSs, respectively. The Voronoi lattice marks the approximate reception area of each BS.}
			\label{fig:5}
		\end{center}
	\end{figure}

	There have been some discoveries on fundamental statistical properties of human mobility\cite{gonzalez2008understanding}, \cite{rhee2011levy}, such as heavy-tail flights and pause-times, heterogeneously bounded mobility areas, and etc. Based on these properties, a number of mobility models have been proposed to generate similar trajectories in real life\cite{privalov2016hybrid, lee2012slaw, munjal2011smooth}. Taking comprehensive consideration of both practicality and complexity of these models, we refer to the Self-Similar Least-Action Human Walk (SLAW)\cite{lee2012slaw} and the SMOOTH model\cite{munjal2011smooth} to generate our mobility data. Specifically, we generate exclusive mobility pattern for each user and capture their location for 12 hours at one-minute granularity in the simulation area of 4000m*4000m each day. At the same time, we consider a single-tier downlink cellular network where the BSs are distributed according to the Poisson Point Process with a density $\lambda$. Fig. \ref{fig:5} shows the simulation area and the trajectory of a single user. For the sake of simplicity, we use BPSK for modulation and demodulation. In addition, the channel bandwidth, the noise power, the path-loss exponent, and the transmission power of each BS is 10 MHz, $10^{-13}$ W, $\alpha=4$, 250 mW, respectively.
	\subsection{Mobility Prediction Results}
	 We use the three-layer LSTM network discussed in Section III to build exclusive prediction model for each user. We take the user's two-dimensional position coordinates of 720 minutes as input and push the time-series forward one minute as standard output. Then the user's position at the next minute can be predicted by the trained model with current and historical position data. The prediction result and the spatial distribution of BSs will be utilized to determine whether an HO is needed in the next minute. We measure the LSTM performance by the HO prediction accuracy which is defined as the ratio of the number of accurately predicted HOs to the number of actual HOs. In order to compare the prediction performance at different speeds, we consider the following three types of UEs: low-speed with 1m/s, medium-speed with 4m/s, and high-speed with 8m/s. We train different LSTM models for nine users, every three of them have the same speed.
	\begin{figure}[htbp]
		\begin{center}
			\includegraphics[width=0.41\textwidth]{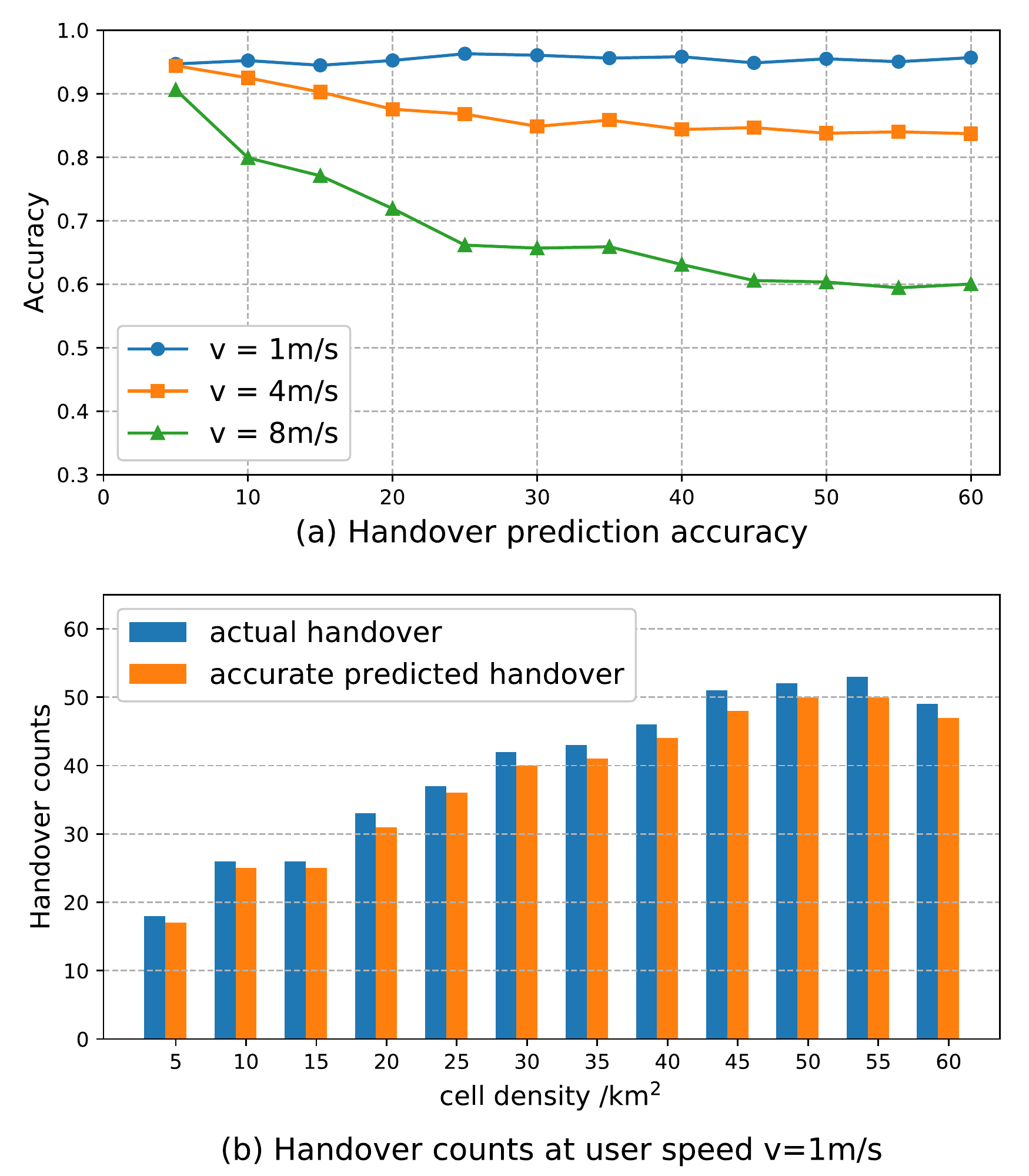}
			\setlength\abovecaptionskip{5pt}
			\setlength\belowcaptionskip{-5pt}
			\captionsetup{font={small}}
			\caption{Handover prediction results}
			\label{fig:6}
		\end{center}
	\end{figure}

	The average handover prediction accuracy of different speed is shown in Fig. \ref{fig:6}(a). It can be observed from Fig. \ref{fig:6}(a), with the increase of cell density, the average handover prediction accuracy of low-speed users always maintains at about 0.95, while the accuracy of the other groups decreases, especially for the high-speed group. This is consist with our intuition, since high speed inevitably leads to mutations in the position data at one-minute granularity. These mutations may result in some deviation in prediction, making it more difficult to predict handover when the coverage of cell is decreasing. Thus, we believe there should be a special mechanism to guarantee the handover of high-speed users, which will be learned as the future work. The following simulation will only consider the case of v=1m/s. Fig. \ref{fig:6}(b) further shows the specific number of actual HOs and accurate predicted HOs of the user with v=1m/s at different cell densities, and indicates that LSTM can predict most of the actual HOs.
	
	\subsection{Performance}
	In this section, we present our simulation results to validate the advantage of our proposed intelligent dual connectivity over conventional handover. Therefore, we only consider the actual handover process in the trajectory data up to 720 minutes. For comparison purposes, we also take account of the ideal dual connectivity, which implies each handover can be predicted successfully.
	
	\begin{figure}[htbp]
		\begin{center}
			\includegraphics[width=0.40\textwidth]{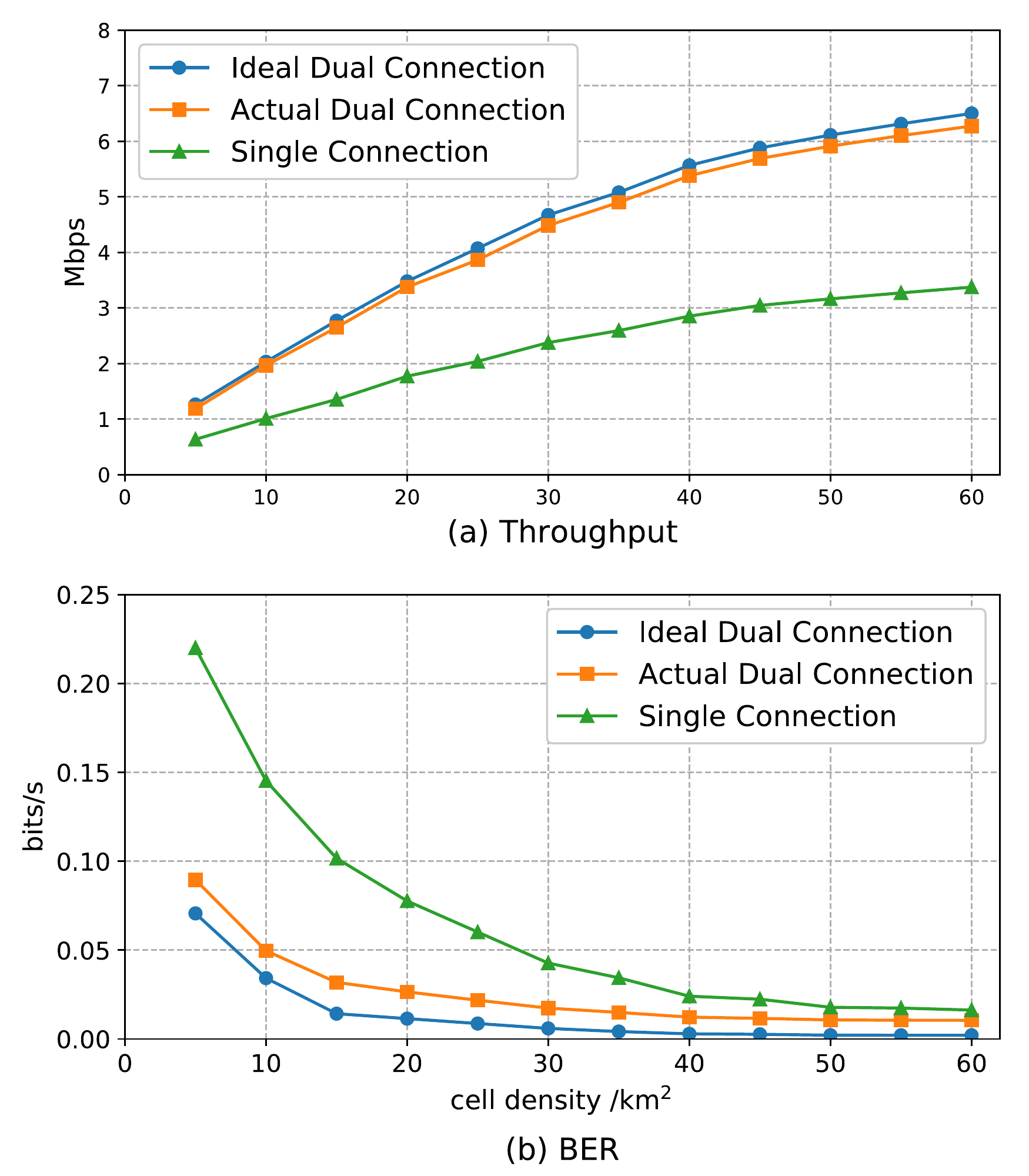}
			\setlength\abovecaptionskip{5pt}
			\setlength\belowcaptionskip{-5pt}
			\captionsetup{font={small}}
			\caption{Average throughput and BER in the HO process (v = 1m/s)}
			\label{fig:7}
		\end{center}
	\end{figure}
	
	Fig. \ref{fig:7}(a) presents the average throughput of the UE in the HO process. It can be observed that at the same cell density, the actual dual connectivity achieves much higher throughput than the single connectivity, and slightly less than the ideal dual connectivity. This is because, during the HO process, the RSS from the serving BS will gradually decrease as the UE moves away from it. However, the UE can get increasing signal from its target BS at the same time if the dual connectivity has been established successfully in advance. 	
	\begin{figure}[htbp]
		\begin{center}
			\includegraphics[width=0.42\textwidth]{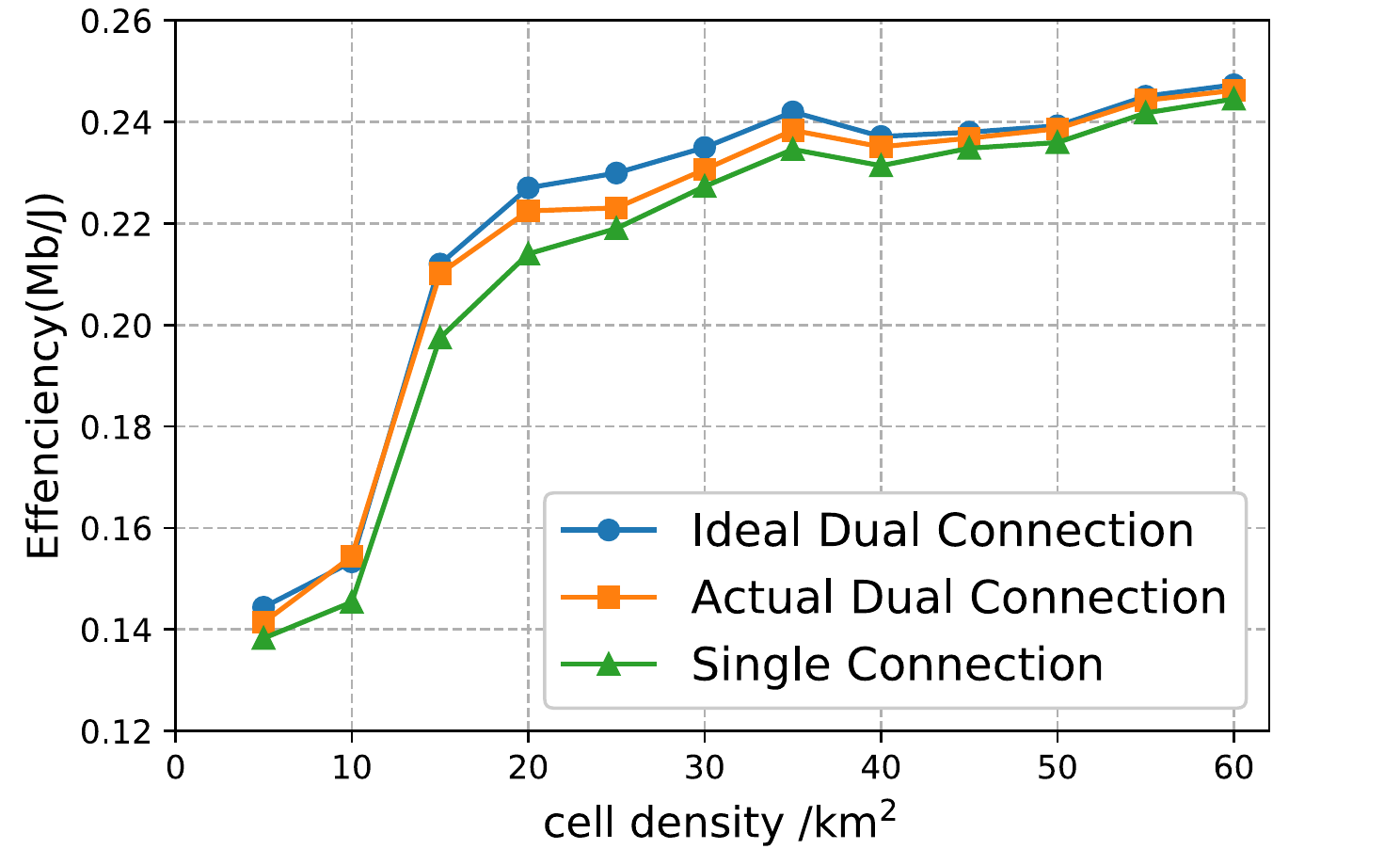}
			\setlength\abovecaptionskip{5pt}
			\setlength\belowcaptionskip{-5pt}
			\captionsetup{font={small}}
			\caption{Energy efficiency in the HO process (v = 1m/s)}
			\label{fig:8}
		\end{center}
	\end{figure}
	
	The bit error rate (BER) of our proposed mechanism is shown in Fig. \ref{fig:7}(b). For a dual-connected UE, we assume its BER is the product of the BERs of both attached BSs, since the error happens only when the signals for both connected BSs are wrong. As expected, the actual dual connectivity yields significantly lower BER than the single connectivity when the cell density is less than 30/km$^2$. As the cell density increases, the difference in BER gradually becomes more narrow. This is because densely deployed BSs have already made the BER of single connection rather low so that the gain from dual connectivity is not obvious.
	
	In Fig. \ref{fig:8}, we evaluate the performance of network energy efficiency (EE), which is defined as the ratio of traffic rate to the energy consumption. In general, the EE increases first and gradually reaches saturation with the increase of cell density. At the same time, both actual and ideal dual connectivity yield some EE improvement to single connectivity, indicating that the throughput gain from dual connectivity is enough to fully compensate for the extra energy consumption. Thus the proposed dual connectivity mechanism has a positive effect in terms of the energy efficiency.
	\begin{figure}[htbp]
		\begin{center}
			\includegraphics[width=0.41\textwidth]{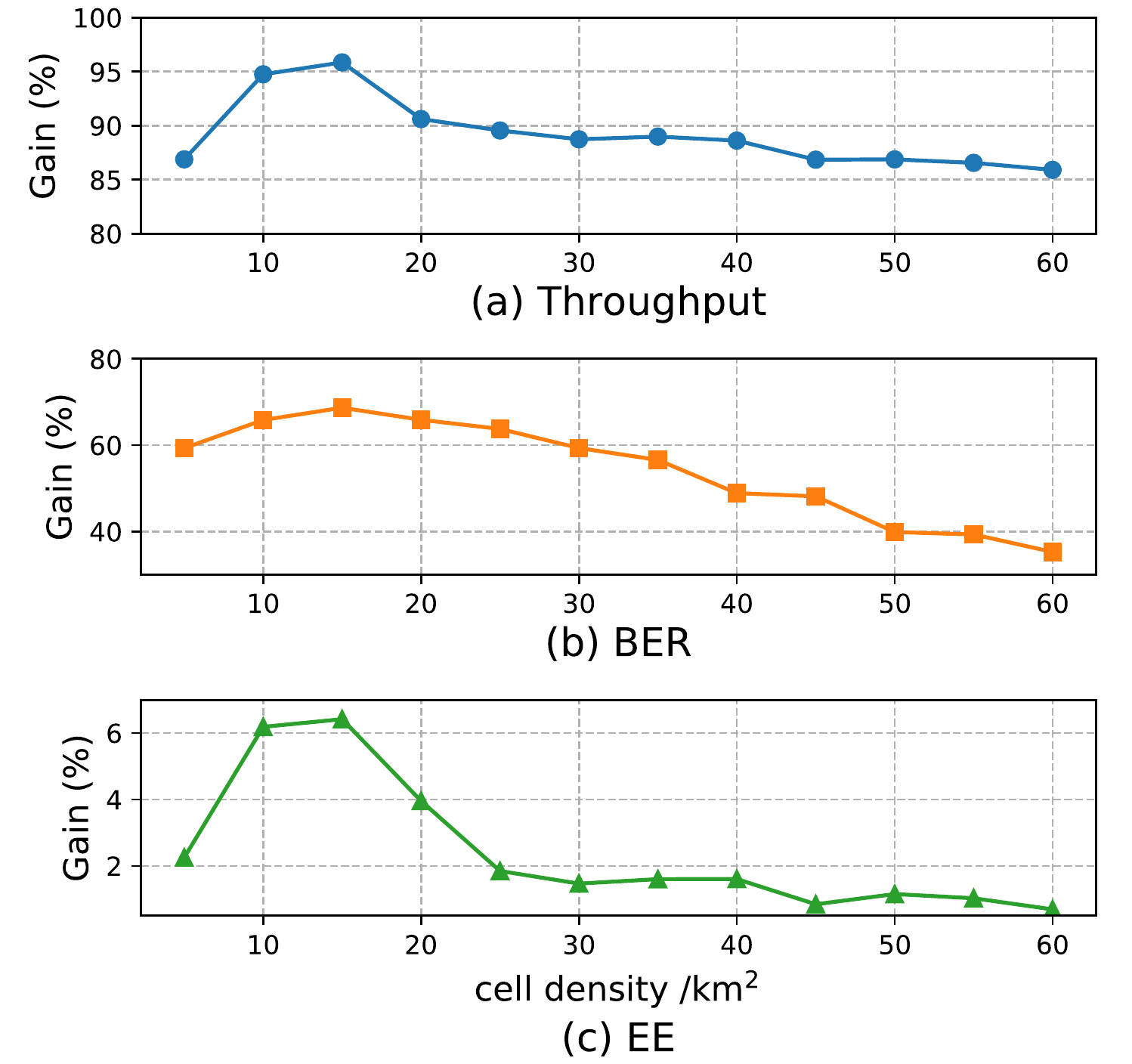}
			\setlength\abovecaptionskip{5pt}
			\setlength\belowcaptionskip{-5pt}
			\captionsetup{font={small}}
			\caption{Comparison of dual connectivity and single connectivity on throughput, bit error rate, and energy efficiency}
			\label{fig:9}
		\end{center}
	\end{figure}
	
	Fig. \ref{fig:9} intuitively shows the gains of dual connectivity in terms of throughput, BER, and EE. It can be observed that the gains increase first and then decrease as the cell density increases. This is because the high cell density environment has already achieved good performance, thus making the dual connectivity gains not obvious. For a dual-connected UE, it can get at least 85\% throughput gain and up to 40\% BER gain in most cases. Meanwhile, the energy efficiency is also slightly improved, especially when the cell density is between 10/km$^2$ and 15/km$^2$. This means the proposed mechanism can bring significant benefits for mobile users in handover process while guaranteeing the network energy efficiency.
	
	\section{Conclusion}
	\label{sec:conclusion}
	In this paper, we have proposed an intelligent dual connectivity mechanism for mobility management in single-tier dense cellular networks. We have used LSTM scheme to learn the user's mobile patterns from its historical trajectories and predicted its movement trends in the future. We have shown high handover prediction accuracy for low-speed and medium-speed users. Based on extensive simulations, we have also verified that the intelligent dual connectivity can bring significant gains in both throughput and bit error rate, thereby improving the QoS of mobile users, while guaranteeing the network energy efficiency. For future work, we will evaluate the performance of our intelligent dual connectivity mechanism on real data. We will also extend the dual-connection to multi-connection and introduce the reinforcement learning to control the connection strategy.
	\bibliographystyle{IEEEtran}
	\bibliography{references}
\end{document}